\documentclass[letterpaper, 10 pt, conference]{ieeeconf}  
\usepackage{amsmath}
\usepackage{algorithm}
\usepackage[noend]{algpseudocode}
\usepackage{algorithm}
\usepackage{algpseudocode}
\usepackage{pifont}
\usepackage{amsfonts}
\usepackage{subcaption}
\usepackage{kantlipsum}       
\usepackage{mwe}  
\usepackage{float}                 
\usepackage{graphicx}
\usepackage{caption}
\usepackage{subcaption}
\usepackage{textcomp}
\usepackage[english]{babel}
\usepackage{multicol}
\usepackage[noadjust]{cite}
\usepackage{textcomp}
\usepackage{mathtools}

\usepackage{hyperref}
\hypersetup{
    colorlinks=true,
    linkcolor=blue,
    filecolor=magenta,      
    urlcolor=cyan,
}

\usepackage[noadjust]{cite}                                                        

\IEEEoverridecommandlockouts                              
\title{\LARGE \bf
Waypoint Optimization Using Bayesian Optimization: A Case Study in Airborne Wind Energy Systems
}

\author{Ali Baheri and Chris Vermillion
\thanks{Ali Baheri is with the Department of Aerospace and Mechanical Engineering, West Virginia University, Morgantown, WV 26505 USA. {\tt\small ali.baheri@mail.wvu.edu}}
\thanks{Chris Vermillion with the Department of Mechanical and Aerospace Engineering, North Carolina State University, Raleigh, NC 27695 USA {\tt\small cvermil@ncsu.edu}.
He is also a technical advisor and equity stakeholder with Altaeros, Inc. and Windlift, LLC. }
}

\begin{document}

\maketitle
\thispagestyle{empty}
\pagestyle{empty}

\begin{abstract}

We present a data-driven optimization framework that aims to address online adaptation of the flight path shape for an airborne wind energy system (AWE) that follows a repetitive path to generate power. Specifically, Bayesian optimization, which is a data-driven algorithm for finding the optimum of an unknown objective function, is utilized to solve the waypoint adaptation. To form a computationally efficient optimization framework, we describe each figure-$8$ flight via a compact set of parameters, termed as basis parameters. We model the underlying objective function by a Gaussian Process (GP). Bayesian optimization utilizes the predictive uncertainty information from the GP to determine the best subsequent basis parameters. Once a path is generated using Bayesian optimization, a path following mechanism is used to track the generated figure-$8$ flight. The proposed framework is validated on a simplified $2$-dimensional model that mimics the key behaviors of a $3$-dimensional AWE system. We demonstrate the capability of the proposed framework in a simulation environment for a simplified $2$-dimensional AWE system model.

\end{abstract}

\section{INTRODUCTION}

Airborne wind energy (AWE) systems are a new paradigm for wind turbines in which the structural elements of conventional wind turbines are replaced with tethers and a lifting body to harvest wind power from significantly increased altitudes where the winds are stronger and more consistent than at ground-level.

\begin{figure}[t]
    \centering
    \includegraphics[width=0.4\textwidth]{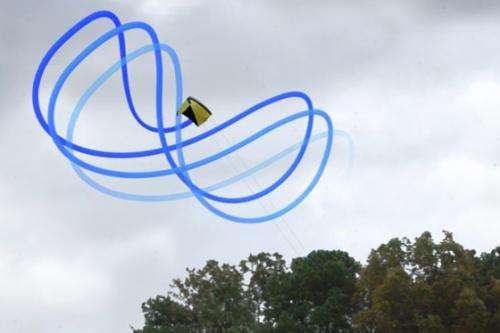}
    \caption{An example of a tethered kite flying in a figure-$8$ to power a generator on the ground. Image credit: \cite{nasa_figure8}}
    \label{fig:figure8ETH}
\end{figure}

Besides being able to operate at much higher altitudes than traditional turbines, AWE systems also provide additional control degrees of freedom that allow the system to intentionally induce \emph{crosswind motions} to enhance power output. Fig. \ref{fig:figure8ETH} shows an example of an AWE system that has been developed to execute repeated crosswind motion, with the objective of increasing the average power output in the long run.

A significant amount of effort in the area of AWE systems has been given to the modeling of AWE systems \cite{vermillion2012modeling}, crosswind motion control \cite{canale2010high,williams2007modeling}, and adaptive control strategies for AWE systems \cite{fagiano2012optimization,isaacs2011retrospective,zgraggen2014real,kehs2017online}. Each of these works aims to optimize or adapt online some set of parameters related to the crosswind motion of the AWE system. Comparatively, fewer studies have focused on directly adjusting the parameters that describe the crosswind flight path, with the ultimate goal of maximizing a net power-based performance index \cite{cobb2017iterative}. In \cite{cobb2017iterative}, the authors first introduce a simplified $2$-dimensional model that mimics the key behaviors of a $3$-dimensional AWE system. Then, motivated from iterative learning control literature, they utilize a learning update rule to adjust the path parameters for a repetitive crosswind flight, with the ultimate goal of maximizing the lap-averaged power.

While the results of \cite{cobb2017iterative} demonstrate effective convergence to the optimal crosswind path (along with a significant associated power generation increase), the results also require a substantial number of figure-8 cycles for convergence. While the convergence rate seen in iterative learning algorithms is sufficient for most wind environments, highly volatile wind environments will require faster convergence than what is achieved by \cite{cobb2017iterative}. The present work builds upon \cite{cobb2017iterative}, where, in contrast to \cite{cobb2017iterative}, we propose a purely data-driven optimization algorithm, namely Bayesian optimization, to adjust a compact set of parameters, termed basis parameters, that fully describe a repetitive flight path subject to maximizing a performance index.

Bayesian optimization has been applied to various real-world problems \cite{baheri2017real,baheri-acc17, baheri2017iterative,baheri_CDBO,baheri2017combined,baheri2018iterative,garnett2010bayesian} as an efficient tool for identifying global optima of complex and non-convex functions. Bayesian optimization intends to find the global optimum of a black-box function within only a few function evaluations. Because of the fast convergence that can be realized, Bayesian Optimization represents an attractive approach for online adaptation in the presence of an environment that is changing rapidly. One popular approach to Bayesian optimization is to model the underlying function as a GP, where it puts prior belief on an objective function to describe the overall structure of that function. At every step of the iterative process, the next operating point (in our case, the next set of basis parameters that describe a flight path) is selected in order to maximize an \emph{acquisition function}, which characterizes (i) how much will be learned by visiting a candidate point (exploration) and (ii) what the likely performance level will be at that next candidate point (exploitation) \cite{brochu2010tutorial}.

The major contributions of this paper can be summarized as follows:

\begin{enumerate}

\item We propose a framework for utilizing Bayesian optimization to adapt waypoint parameters of a repetitive crosswind flight path, with the goal of maximizing a performance index.

\item We illustrate the capability of the proposed framework for waypoint adaptation for a simplified $2$-dimensional AWE system model, along with the comparatively fast convergence of path parameters to their optimal values.

\end{enumerate}

The rest of this paper is structured as follows: In Section \ref{sec:model}, we present the simplified $2$-dimensional model considered in this work, the wind profile model, and the performance index. Next, Section \ref{sec:method} formalizes the basis parameterization technique and Bayesian optimization algorithm for waypoint adaptation. Finally, Section \ref{sec:results} provides detailed results, demonstrating the effectiveness of the proposed approach.




\section{$2$-dimensional sailboat, wind model, and performance metric}
\label{sec:model}

\subsection{$2$-dimensional sailboat}

To validate our approach for an AWE application, we use a $2$-dimensional model developed in \cite{cobb2017iterative} to mimic the key behaviors of a $3$-dimensional AWE system. This $2$-dimensional, termed the \emph{sailboat model}, approximately characterizes the planar projection of the AWE system\textquotesingle s motion. The model consists of a hull and a sail. The direction of the hull is controlled by a rudder that dictates the direction of motion. 

The orientation of the hull, $\psi$, evolves based on the following dynamic model:

\begin{equation}
\dot{r} = \frac{k_r}{I_R}v^2u_r,
\end{equation}
\begin{equation}
\dot{\psi} = r,
\end{equation}
for $ -\frac{\pi}{2} \leq u_r \leq \frac{\pi}{2}$. Here, $r$ represents the hull\textquotesingle s angular velocity, $v$ represents
the speed of the sailboat, and $u_r$ represents the rudder angle. Furthermore, the constant parameters $k_r$ and $I_R$ are a lumped rotational aerodynamic damping coefficient and the hull\textquotesingle s inertia, respectively.

We model the lift and drag forces, $F_L$ and $F_D$, respectively, which depend upon the angle of attack, $\alpha$, as follows:

\begin{equation}
F_L(\alpha) = k_L\alpha v_{\mathrm{app}}^2,
\end{equation}
\begin{equation}
F_D (\alpha)= \big(k_{D0} + k_{D1}\alpha^2\big) v_{\mathrm{app}}^2,
\end{equation}
where $\alpha$ represents the angle between the apparent wind vector, $v_\mathrm{app}$, and the sailboat. Specifically, $\alpha = \psi_{\mathrm{app}} - u_s$. Furthermore, $k_L$, $k_{D0}$, and $k_{D1}$ represent the lumped lift sensitivity, lumped drag coefficient at $\alpha = 0$, and lumped drag sensitivity, respectively.

\begin{figure}[t]
    \centering
    \includegraphics[width=0.3\textwidth]{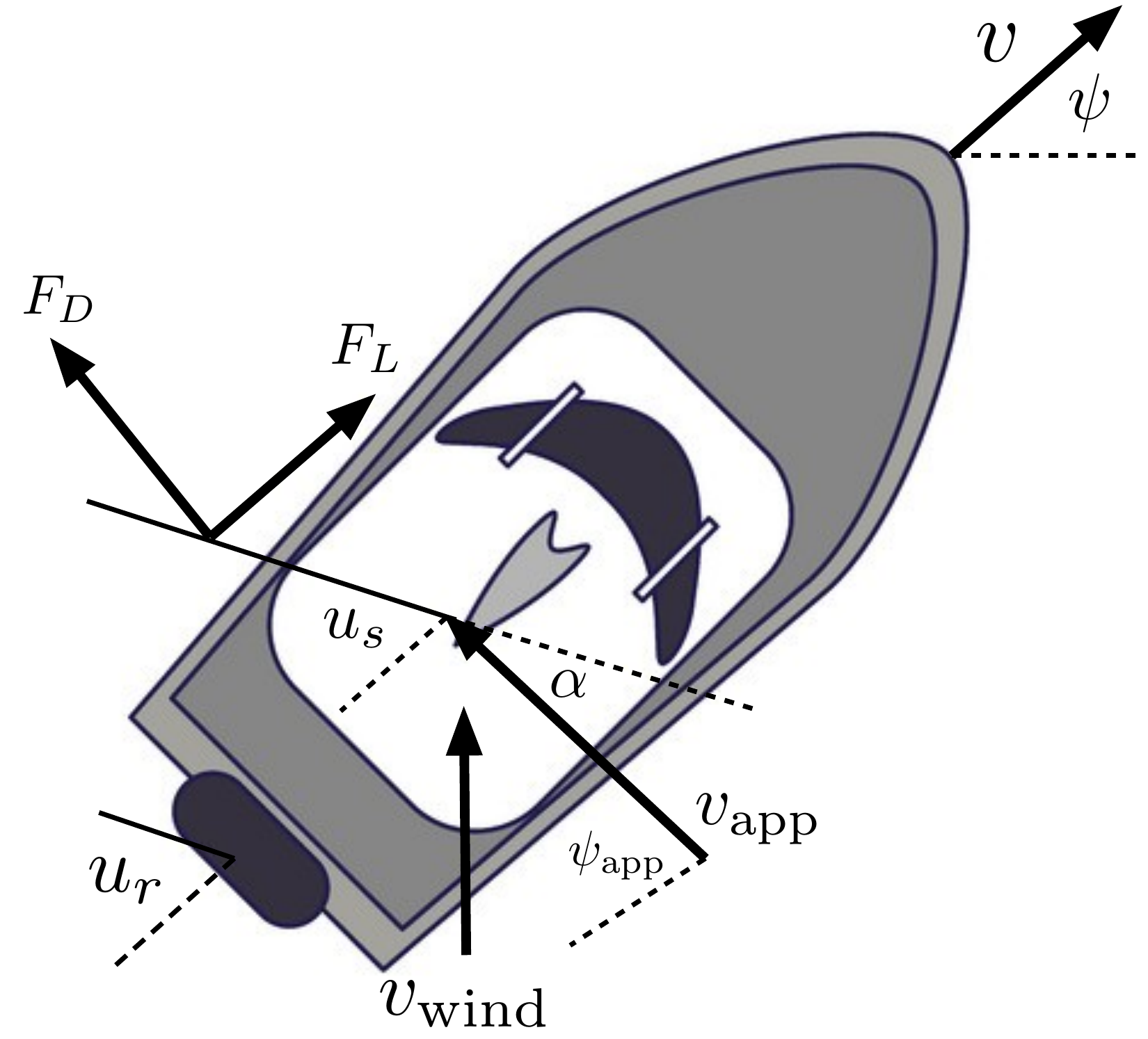}
    \caption{A simplified $2$-dimensional sailboat model. We use a simplified $2$-dimensional  model that mimics the behaviors of a $3$-dimensional AWE system. This figure shows various angles and forces presented at Section \ref{sec:model}. Adapted from \cite{cobb2017iterative}.}\    
    \label{fig:sailboat}
\end{figure}
\begin{figure}[t]
    \centering
    \includegraphics[width=0.5\textwidth]{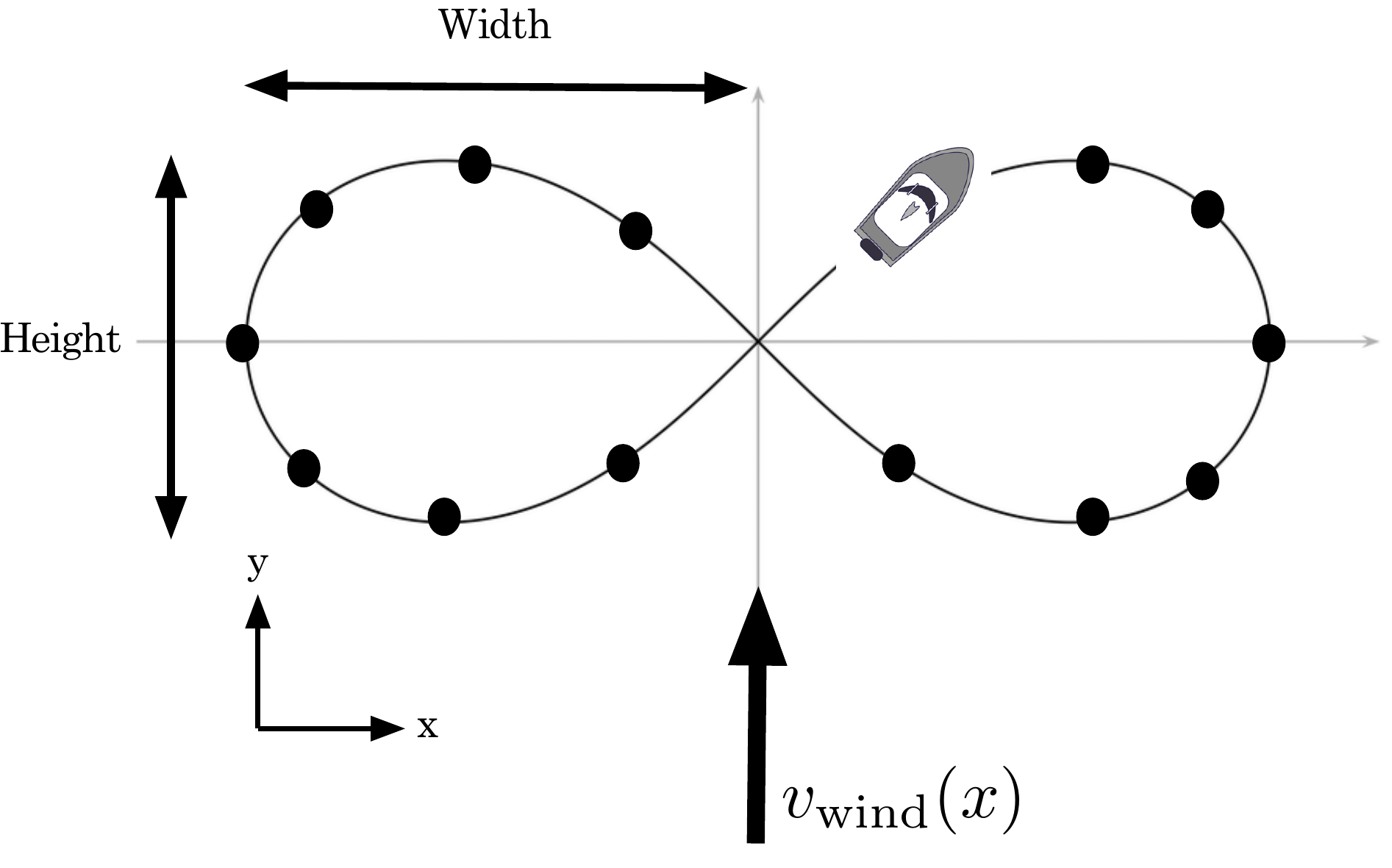}
    \caption{Each figure-$8$ path can be parameterized using several waypoints that ensure us a close approximation to a smooth figure-$8$ path flight. A simple wind profile is used parallel to the plan of sailboat.}
    \label{fig:figure8}
\end{figure}

The hull velocity vector is aligned with its heading $\psi$. Therefore, the translational equations of motion are as follows:
\begin{equation}
\dot{v} = \frac{1}{m}\big(F_L (\alpha) \sin(\psi_{\mathrm{app}}) - F_D (\alpha),\cos(\psi_{\mathrm{app}})\big),
\end{equation}
\begin{equation}
\dot{x} = v \cos(\psi),
\end{equation}
\begin{equation}
\dot{y} = v \sin (\psi).
\end{equation}
Here, the apparent wind heading, $\psi_{\mathrm{app}}$, is measured relative
to the system heading, $\psi$, in the two dimensional plane. Fig. \ref{fig:sailboat} demonstrates the key angles of the simplified sailboat model.

\subsection{Wind Model}

In this work, we consider a simple wind velocity profile, $v_{\mathrm{wind}}(x)$, as follows:

\begin{equation}
    v_{\mathrm{wind}}(x) = 
\begin{cases}
    v_{\mathrm{max}} \cos (\frac{\pi x}{2R}) ,& \text{if} -R \leq x \leq R,\\
    0,              & \text{otherwise.}
\end{cases}
\end{equation}
\noindent This model captures the fact that the available wind is greatest at the center of the so-called \emph{wind window} in the crosswind flight of AWE systems, and dissipates as the azimuthal position of the AWE system is increased (manifested in the sailboat model by an increase in the magnitude of $x$).

\subsection{Performance Metric}

The ultimate objective considered in this paper is to maximize power output in a repetitive manner from one iteration to the next. To achieve this goal, and recognizing that the power produced by an on-board turbine on an AWE system is proportional to the apparent wind speed \emph{cubed}, the following performance index, $J$, can be chosen as follows:
\begin{equation}
J  = \frac{k_p}{T_f}\int_{0}^{T_f}v_{\mathrm{app}}^3(t)dt,
\end{equation}
where $k_p$ and $T_f$  represent a lumped power coefficient and the time required to complete the crosswind flight, respectively.

\section{Methodology}
\label{sec:method}

The ultimate goal with Bayesian optimization is to maximize the average power output generated by adapting the waypoint parameters. In this section, we formalize a computationally efficient optimization framework for waypoint adaption and fundamental of Bayesian optimization. Once the waypoints are generated, a path following mechanism is introduced to track these waypoints.

\section{Waypoint parameterization}

To form a computationally efficient optimization framework, we describe each figure-$8$ flight via two \emph{basis parameters}. We parameterize each figure-8 path using $40$ waypoints, which achieve a close approximation to a smooth figure-$8$ (See Fig. \ref{fig:figure8}).

The basis parameters will be defined in this work through a compact vector, $\beta$, where:

\begin{equation}
\beta :=  \begin{bmatrix}
W & H 
\end{bmatrix}^T,
\end{equation}
where $W$ and $H$ represent the weight and height of a figure-$8$ flight path. Each waypoint, $(x_i,y_i)$, is computed from a parametric form of the figure-$8$ known as the Lemniscate of Gerono \cite{delgado2007progressive}, where:

\begin{equation}
x_i = W \cos (\frac{2\pi i}{n+1}),
\end{equation}
\begin{equation}
y_i = H \sin(\frac{2\pi i}{n+1}) \cos(\frac{2\pi i}{n+1}).
\end{equation}

\subsection{Bayesian optimization}

Consider the goal of maximizing an unknown function that may contain multiple local or corner extrema. Due to the computational costs associated with evaluating this function and the possibility of the optimization getting stuck in a local optimum, it is important to select the location of each new evaluation deliberately. In such cases, Bayesian optimization can be used to find the \emph{global optimum} of the objective function within \emph{few evaluations} on the real system.

Bayesian optimization consists of two steps. First, at each iteration, we update a model that characterizes our best guess at the objective function vs. basis parameters, along with a characterization of the uncertainty in that guess. We refer this phase as of Bayesian optimization as the \emph{learning phase}. Second, we choose an \emph{acquisition function}, which guides the optimization by determining the next basis parameters to evaluate. The selection of the next basis parameters in an effort to maximize the acquisition function is referred to as the \emph{optimization phase}.

\subsubsection{\textbf{Learning Phase: Using Gaussian Processes (GPs)}}
\label{GP}
In this section, we introduce the basic properties of Gaussian Process (GP) models. GP models will be used in this work to characterize, at each iteration, our best guess of average power output vs. basis parameters, along with the associated uncertainty in this guess. 

As an attractive choice for non-parametric regression in machine learning, the ultimate goal of GP models is to find an approximation of a nonlinear map, from an input vector to the objective function value. GP models are able to model complex phenomena since they are able to handle the nonlinear effects and interaction between covariates in a systematic fashion. GP models assume that the objective function values ($J$ in our case) associated with different inputs (i.e., basis parameters, as specified through the basis parameter vector $\beta$ in our case), have joint Gaussian distributions \cite{rasmussen2006gaussian}. While the true average power output is deterministic, not stochastic, the Gaussian model of the average power output allows us to treat modeling uncertainty through the concept of variance.

A GP model is fully specified by its mean function, $\mu(\beta)$, which is assumed to be zero without loss of generality \cite{rasmussen2006gaussian} (a coordinate shift can be used to generate a zero mean objective function, as is done in our case where $J$ is our ultimate objective function), and covariance function, $k(\beta,\beta\textquotesingle)$:
\begin{equation}
J(\beta) \sim \mathcal{GP}\Big(\mu(\beta),k(\beta,\beta\textquotesingle)\Big),
\end{equation}
The GP framework is used to predict the generated average power, $J(\beta)$, at any given basis parameter, $\mathcal{\beta}$, based on a set of $t$ past observations, $\mathcal{D}_{1:t} = \begin{Bmatrix}\beta_{1:t},J(\beta)\end{Bmatrix}$. The function value, $J(\beta^{\ast })$, for an unobserved input $\beta^{\ast }$, and the observed function values, follow a multivariate Gaussian distribution \cite{rasmussen2006gaussian}:

\begin{equation}
\begin{bmatrix}
y_{t}\\ J(\beta^{\ast })\end{bmatrix}  \sim  \mathcal{N}\Big(0,\begin{bmatrix}
K_{t} + \sigma_{\epsilon}^{2}I_{t}&k_{t}\\ 
 k_{t}^{T}& k(\beta,\beta)
\end{bmatrix}\Big),
\label{eqn:multivariate_g}
\end{equation}
where $y_{t} = \begin{Bmatrix}
J(\beta_1), \cdots, J(\beta_t)
\end{Bmatrix}$ is the vector of observed function values. The vector $k_{t}(\beta) = [k(\beta,\beta_{1}), \cdots ,k(\beta,\beta_{t})]$ encodes the relationship between the new input, and the past data points in $\mathcal{\beta}$. The covariance matrix, denoted by $K_t$, has entries $[K_{t}]_{(i,j)} =k(\beta_{i},\beta_{j})$ for $i,j \in \begin{Bmatrix}
1, \cdots, t
\end{Bmatrix}$. Finally, the identity matrix is represented by ${I}_{t}$, and $\sigma_{\epsilon}$ represents the noise variance \cite{rasmussen2006gaussian}. 

Individual elements of the covariance matrix, namely $k(\beta_{i},\beta_{j})$, encode the correlation between two different sets of basis parameters. In order to characterize this correlation in a closed-form manner, a \emph{covariance kernel} is used. This covariance kernel provides a relatively simple parametric structure for the values of $k(\beta_{i},\beta_{j})$. In this study, we represent the elements of the covariance matrix through a commonly used kernel function, known as the Squared Exponential (SE) covariance kernel. For two sets of basis parameters, $\beta_i$ and $\beta_j$, the SE kernel is parameterized as:

\begin{equation}
k(\beta_i,\beta_j) = \sigma_{0}^2 \ {\exp} \Big(-\frac{1}{2}(\beta_i-\beta_j)^T \Lambda^{-2}(\beta_i-\beta_j)\Big).
\end{equation}
Here, $\theta = \begin{Bmatrix}\sigma_{0},\Lambda\end{Bmatrix}$ are the hyper-parameters of the kernel function. We identify these hyper-parameters by maximizing the marginal log-likelihood of $\mathcal{D}$ \cite{rasmussen2006gaussian} over the domain of feasible hyper-parameters, as follows:
 
\begin{equation}
\theta^{\ast} = \underset{\theta}{\arg \max \log}\  p(y_t\mid \mathcal{\beta},\theta),
\label{eqn:hyper-p}
\end{equation}
where

\begin{multline} 
{\log} \ p(y_t\mid \mathcal{\beta},\theta) = \Big(-\frac{1}{2}y_{t}^{T}K^{-1}y_{t}-\frac{1}{2}$log$\mid K\mid-\frac{t}{2}$log$2\pi\Big).
\end{multline} 
Once the hyper-parameters are optimized, the predictive mean and variance at $\beta^{\ast }$, conditioned on these past observations, are expressed as:
\begin{equation}
\mu_{t}(\beta^{\ast}\mid \mathcal{D} ) = k_{t}(\beta)\Big(K_{t}+{I}_{t}\sigma_{\epsilon}^{2}\Big)^{-1}y_{t}^{T},
\label{eqn:gp_mean}
\end{equation}
\begin{equation}
\sigma_{t}^{2}(\beta^{\ast }\mid \mathcal{D}) = k(\beta,\beta) - k_{t}(\beta)\Big(K_{t} + {I}_{t}\sigma_{\epsilon }^{2}\Big)^{-1}k_{t}^{T}(\beta).
\label{eqn:gp_var}
\end{equation}
In short, the learning phase using GPs consists of two main steps: training and prediction. The training step involves finding proper a mean function, a covariance function, and optimized hyper-parameters in the light of the data (Eq. \ref{eqn:hyper-p}). The prediction phase, predicts the objective function value at an unobserved input in a probabilistic framework (Eqs. \ref{eqn:gp_mean}-\ref{eqn:gp_var}) \cite{rasmussen2006gaussian}. These two equations implicitly serves as a surrogate of the unknown function and are used in the next phase to form the acquisition function.


\begin{algorithm}[t]
\caption{}
\label{alg:BO}
\begin{algorithmic}[1]
\Procedure{Waypoint optimization with Bayesian optimization}{}
\State $\mathcal{D}\gets \textit{Initialize}$$: \begin{Bmatrix}\beta_{1:5},J(\beta_{1:5})\end{Bmatrix}$ 
\For{each iteration}
\State Train a GP model from $\mathcal{D}$
\State Compute mean and variance of GP \big(Eq. \ref{eqn:gp_mean}-\ref{eqn:gp_var}\big)
\State Compute acquisition function \big(Eq. \ref{eqn:EI}\big)
\State Find $\beta^{\ast}$ that optimizes acquisition function
\State Append $\begin{Bmatrix}\beta^{\ast},J(\beta^{\ast})\end{Bmatrix}$ to $\mathcal{D}$
\EndFor
\EndProcedure
\end{algorithmic}
\label{alg:BO}
\end{algorithm}


\subsubsection{\textbf{Optimization Phase}}
The optimization phase of Bayesian optimization hinges upon the maximization of an \emph{acquisition function}, which guides the optimization by determining the next basis parameters to evaluate.

Among several choices of acquisition functions, we use an acquisition function known as \emph{expected improvement} \cite{brochu2010tutorial}. Specifically, we choose the basis parameters for the next iteration, denoted by $\beta_{t+1}$, as follows:
\begin{equation}\label{acquisition_function}
\beta_{t+1} =\underset{\beta} {\arg\max}\ \mathbb{E}\Big(\max \begin{Bmatrix}0,J(\beta)-J^{\emph{max}} \end{Bmatrix}\mid \mathcal{D}_{1:t}\Big),
\end{equation}
\noindent where $\max \begin{Bmatrix}0,J(\beta)-J^{\emph{max}} \end{Bmatrix}$ represents the \emph{improvement} toward the best value of the objective function so far, $J^{\emph{max}}$. The improvement is positive when the prediction is higher than the best value of the objective function so far. Otherwise, it is set to zero. The inability for the acquisition function to assume negative values reflects the fact that if the performance \emph{worsens} from one iteration to the next, then it is possible to simply revert to the previous best set of basis parameters. This feature is what incentivizes exploration within the expected improvement framework. The expected improvement (i.e., the right side of Eq. \ref{acquisition_function}) will be denoted by $\mathbb{E}\mathbb{I}$. Expected improvement can be expressed in closed form as in \cite{mockus_ei}: 

\begin{multline} 
\mathbb{E}\mathbb{I} (\beta_{t+1}\mid \mathcal{D}_{1:t}) = \\ \begin{cases}
\Big(\mu_{t}(\beta)-J(\beta)^{\emph{max}}\Big)\Phi(Z) + \sigma_{t}(\beta)\phi(Z),& \sigma_{t}(\beta) > 0 \\ 
0, & \sigma_{t}(\beta)= 0
\end{cases}
\label{eqn:EI}
\end{multline} 
where 
\begin{equation}
Z = \frac{\mu_{t}(\beta)-J(\beta)^{\emph{max}}}{\sigma_{t}(\beta)}.
\end{equation}
Here, $\Phi(.)$ and $\phi(.)$ denote the cumulative and probability density function for the normal distribution, respectively.

\subsection{Path following mechanism}

\begin{figure}[t]
    \centering
    \includegraphics[width=0.5\textwidth]{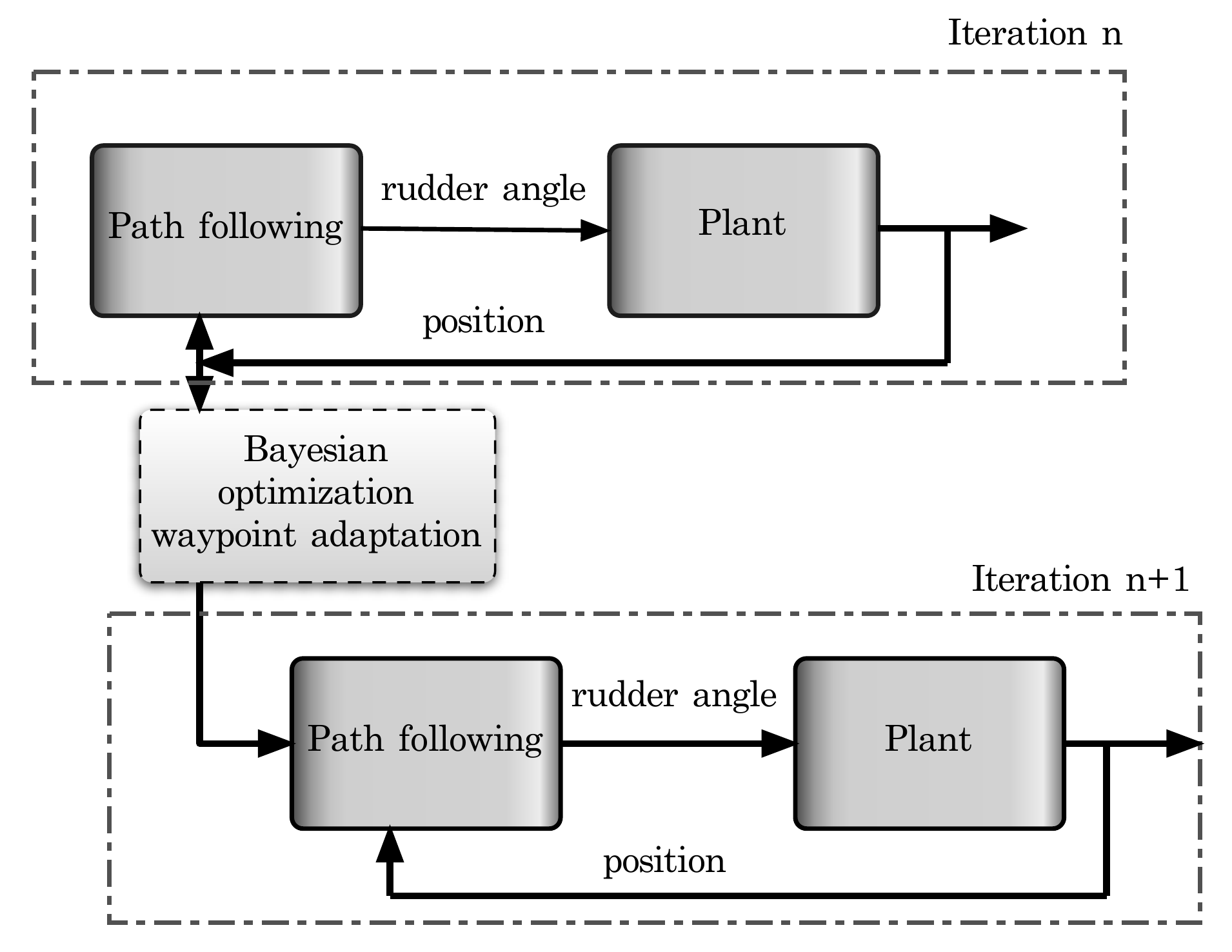}
    \caption{Overall control architecture consisting of (i) Bayesian optimization to generate the optimized shape of figure-$8$ flight path and (ii) a low-level controller to track the optimized waypoints.}
    \label{fig:low_level}
\end{figure}

Once the path is generated by Bayesian optimization, a tracking mechanism is required to follow the generated path. Fig. \ref{fig:low_level} demonstrates the overall proposed control architecture, consisting of waypoint adaptation using Bayesian optimization and a path following mechanism that allows the AWE system to follow the path generated by Bayesian optimization. In this work, the path following mechanism is a feedback linearizing model reference controller that is spelled out in detail in \cite{cobb2017iterative}.

\section{Results}
\label{sec:results}

We evaluate the effectiveness of the proposed framework in simulation for a simplified $2$-dimensional sailboat model. Based on Algorithm \ref{alg:BO}, the Bayesian optimization is initialized with five arbitrary basis parameters and associated average power outputs. Running each iteration of Bayesian optimization leads to a new set of basis parameters, $\beta$. Then, the estimated power output is evaluated using the performance index, $J$, outlined in section \ref{sec:model}. The new data is augmented to the initialized data, and Bayesian optimization identifies the next set of basis parameters. This process continues while the optimal basis parameters are found. 

\begin{figure}[t]
    \centering
    \includegraphics[width=0.5\textwidth]{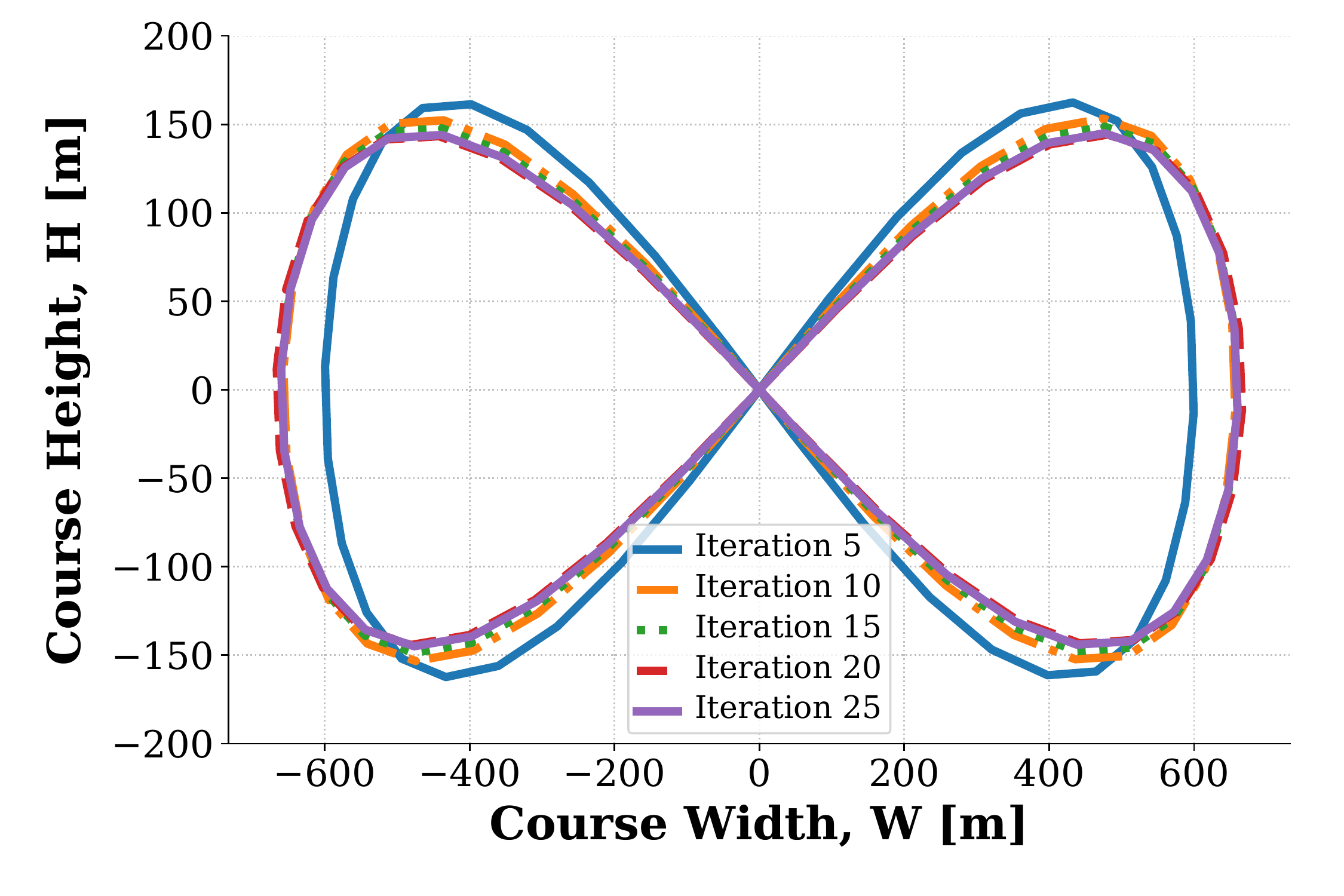}
    \caption{Evolution of figure-8 path for a certain initial condition. One can conclude that the course geometry converges to its optimal shape within a few iterations.}
    \label{fig:waypoint}
\end{figure}

\begin{figure}[t]
    \centering
    \includegraphics[width=0.5\textwidth]{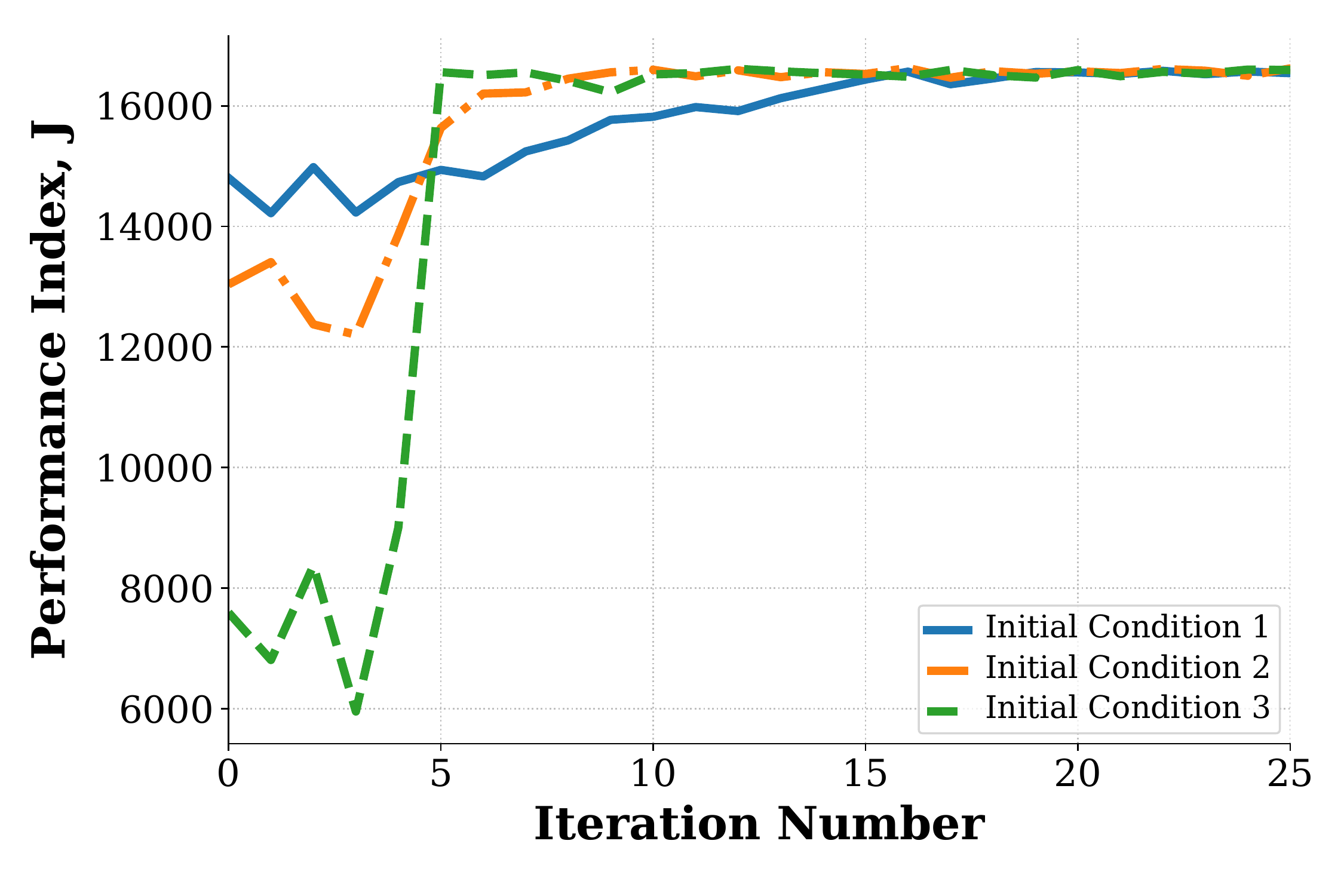}
    \caption{Evolution of performance index for a variety of initial conditions.}
    \label{fig:performance}
\end{figure}

\begin{figure}[t]
    \centering
    \includegraphics[width=0.5\textwidth]{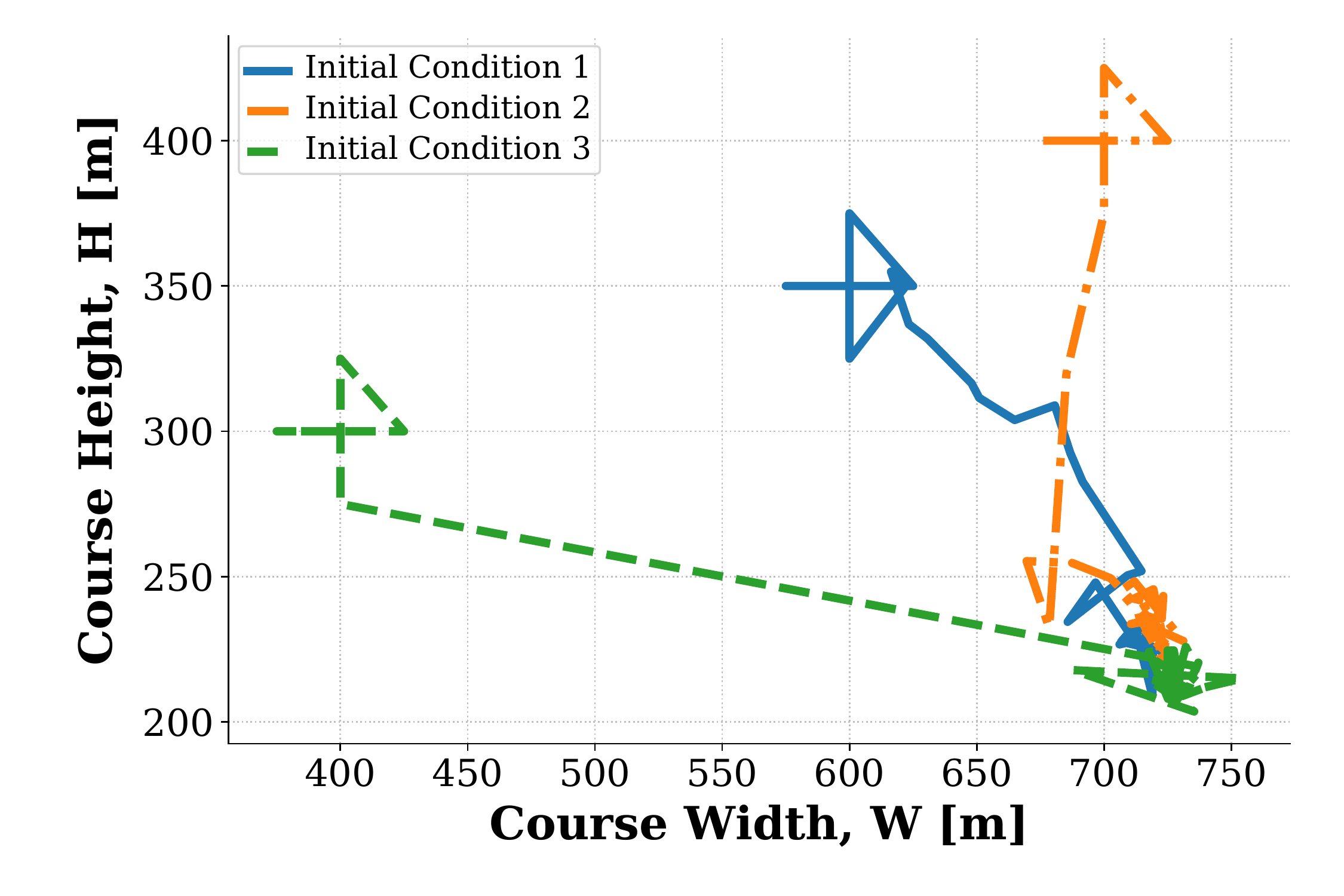}
    \caption{Evolution of basis parameters for a variety of initial conditions on the basis parameters. With a variety of different initial conditions the course geometry basis parameters converge to an approximate optimal value.}
    \label{fig:trajectories}
\end{figure}

Fig. \ref{fig:waypoint} shows the course geometry over a single optimization process at several iterations. This figure demonstrates that the figure-8 path converges as iterations increase. Fig. \ref{fig:performance} shows the evolution of the basis parameters at each iteration, for a variety of initial conditions on basis parameters. With the same simulation settings, comparing these results with those presented in \cite{cobb2017iterative} reveals that the proposed optimization technique leads to a faster convergence rate than the iterative learning approach. It can be seen that for a given initial condition, the course geometry converges to its optimal shape within only a few function evaluations. Fig. \ref{fig:trajectories} shows the evolution of basis parameters for a variety of initial conditions. Based on these results, convergence is robust to the initial figure-8 path geometry.
  
\subsection{Discussion and future directions}

This paper is our first attempt to study the feasibility of Bayesian optimization for repetitive path planning with application to AWE systems. We intend to extend the current framework in several directions. Thus far, we have modeled the wind velocity profile using a simple model. The comparison of the current results with a realistic wind profile is left as future work. Furthermore, the current framework utilized a simplified $2$-dimensional sailboat model. We will consider a high-fidelity model in the future to dynamically capture the full AWE system. Finally, we will perform a regret analysis to study the convergence of the Bayesian optimization algorithm in the presence of a truly stochastic wind environment. 

\section{CONCLUSIONS}

This work presented a data-driven optimization framework that aims to address online adaptation of the course geometry in a repetitive path planning application. To achieve that goal, we utilized Bayesian optimization, a data-driven optimization algorithm, to find the optimum of an unknown objective function. We validated the proposed framework on a simplified $2$-dimensional model of an AWE system, which serves as an analogy to the more complex $3$-dimensional system. To form a computationally efficient optimization framework, we described each figure-$8$ flight path via a compact set of parameters, termed as basis parameters. The performance metric was then modeled by a GP, where the Bayesian optimization utilized the predictive uncertainty information from the GP to determine the best subsequent basis parameters. Compared to a study in \cite{cobb2017iterative}, which used iterative learning techniques on the same application, we demonstrated that the basis parameters converge to optimal values within a fewer objective function evaluations.

\bibliographystyle{unsrt}
\bibliography{Bo_Summer2016}

\end{document}